\documentclass[final,times,twocolumn]{elsarticle}
\usepackage{graphics}
\usepackage{amssymb}

\journal{Physics Letters A}

\begin{document}

\begin{frontmatter}



\title{Spin-phonon coupling induced frustration in the exactly solved 
spin-1/2 Ising model on a decorated planar lattice\tnoteref{fapemig}}       
\tnotetext[fapemig]{The financial support of this work provided by the research foundation FAPEMIG 
under the grant number BPV-00088-10 is gratefully acknowledged.}
\author[upjs]{Jozef Stre\v{c}ka\corref{cor1}}
\cortext[cor1]{J.S. acknowledges stimulating discussions with V. Ohanyam 
who brought his attention to spin models with a magnetoelastic coupling.}
\ead{jozef.strecka@upjs.sk}
\ead[url]{http://158.197.33.91/$\thicksim$strecka}
\author[ufla]{Onofre Rojas} 
\author[ufla]{S.M. de Souza} 
\address[upjs]{Department of Theoretical Physics and Astrophysics, 
Faculty of Science, P. J. \v{S}af\'{a}rik University, Park Angelinum 9,
040 01 Ko\v{s}ice, Slovakia}
\address[ufla]{Departamento de Ciencias Exatas, Universidade Federal de Lavras, 
CP 3037, 37200000, MG, Brazil}

\begin{abstract}
The spin-1/2 Ising model with a spin-phonon coupling on decorated planar lattices partially amenable 
to lattice vibrations is examined within the framework of the generalized decoration-iteration 
transformation and the harmonic approximation. It is shown that the magnetoelastic coupling gives rise 
to an effective antiferromagnetic next-nearest-neighbour interaction, which competes with the nearest-neighbour interaction and is responsible for a frustration of the decorating spins.
The strong enough spin-phonon coupling consequently leads to an appearance of the striking partially
ordered and partially disordered phase, where a perfect antiferromagnetic alignment of the nodal 
spins is accompanied with a complete disorder of the decorating spins. The diversity
in temperature dependences of the total specific heat is investigated in connection with the particular behaviour of its magnetic and lattice contribution.
\end{abstract}


\end{frontmatter}


\section{Introduction}
\label{intro}

Critical properties of compressible spin systems are traditionally subject of extensive theoretical studies, since they depend in a rather crucial and subtle way on specific constraints laid on a given compressible 
spin model. A vast number of theoretical studies on the compressible spin-1/2 Ising planar 
models serves in evidence that the magnetoelastic coupling may either change a continuous (second-order)  phase transition to a discontinuous (first-order) one \cite{rice54,domb56,bean62,matt63,lark69,sali73,sali74}, 
cause a renormalization of critical exponents in a continuous phase transition \cite{bake70,bake71,gunt71,berg73,pens73}, or lead to a striking cross-over between both aforedescribed 
scenarios at a tricritical point \cite{bergi73,imry74,dohm74,ried74,berg76,cape78,cape79,oude81}. From this point of view, it is quite valuable to investigate this rather subtle inter-relationship between the critical behaviour and specific contraints laid on compressible spin systems especially with the help of 
exactly tractable spin models.  

In the present Letter, we will introduce a new class of exactly soluble Ising models on decorated planar lattices (see Ref. \cite{syoz72} and refernces cited therein) in which atoms from nodal lattice 
sites are excluded from lattice vibrations in contrast with decorating atoms that are relaxed from 
a condition of perfect lattice rigidity. The lattice vibrations of the decorating atoms will be treated 
as quantum harmonic oscillations and will consequently lead to a presence of the competing next-nearest-neighbour interaction, which basically affects the critical behaviour of the investigated spin system whenever a relative strength of the magnetoelastic interaction exceeds some threshold value. However, our assumption on a lattice rigidity of the nodal atoms is sufficient to prevent a change in the character of a phase transition, which always turns out to be from the standard Ising universality class.
 
This Letter is so organized. The investigated model is defined in Section \ref{model}, where 
the basic steps of the calculation procedure are also explained. The most interesting results 
for the ground-state and finite-temperature phase diagrams, the spontaneous magnetization and 
the specific heat are presented in Section \ref{result}. Finally, some conclusions and future outlooks 
are presented in Section \ref{conclusion}.

\section{Ising model and its exact solution}
\label{model}

\begin{figure}
\begin{center}
\includegraphics[width=0.3\textwidth]{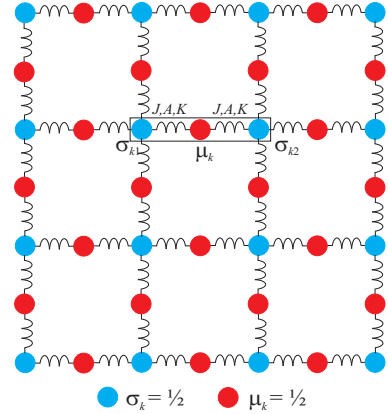}
\end{center}
\vspace{-0.6cm}
\caption{The part of a decorated square lattice. Blue circles label rigid lattice positions of the nodal 
atoms with the Ising spin $\sigma=1/2$, while the red ones denote lattice positions of the decorating  
atoms with the Ising spin $\mu=1/2$ that are relaxed from the condition of a perfect rigidity.}
\label{fig1}       
\end{figure}

Let us begin by considering the spin-1/2 Ising model on decorated planar lattices such as a decorated 
square lattice schematically depicted in Fig.~\ref{fig1}. The decorated lattice constitute two inequivalent lattice sites to be further referred to as nodal and decorating sites, respectively, which are occupied 
by the nodal atoms with the Ising spin $\sigma=1/2$ and the decorating atoms with the Ising spin $\mu=1/2$. Suppose furthermore that the nodal atoms are placed at rigid lattice positions contrary to the decorating atoms, which are relaxed from the condition of a perfect rigidity. Under these circumstances, it is advisable to define the total Hamiltonian of the investigated 
model system as a sum of bond Hamiltonians, whereas each bond Hamiltonian $\hat{{\cal H}}_k$ 
contains all the interaction terms of the $k$th decorating atom and is further divided into 
the magnetoelastic part $\hat{{\cal H}}_{k,m}$ and the pure elastic part $\hat{{\cal H}}_{k,l}$ 
\begin{equation}
\hat{\cal H} = \sum_{k=1}^{Nq/2} \hat{\cal H}_k = \sum_{k=1}^{Nq/2} (\hat{\cal H}_{k,m} + \hat{\cal H}_{k,l}).
\label{ha}
\end{equation}
Above, $N$ denotes the total number of the nodal atoms and $q$ determines the number of their nearest neighbours. The magnetoelastic part of the bond Hamiltonian $\hat{\cal H}_{k,m}$ takes into account the 
exchange interaction, which depends on an instantaneous distance between the decorating spin $\mu_k$ 
and its two nearest-neighbour nodal spins $\sigma_{k1}$ and $\sigma_{k2}$
\begin{eqnarray}
\hat{{\cal H}}_{k,m} = - (J - A \hat{x}_{k}) {\hat{\mu}}_{k}^z {\hat{\sigma}}_{k1}^z 
                     - (J + A \hat{x}_{k}) {\hat{\mu}}_{k}^z {\hat{\sigma}}_{k2}^z.
\label{ham}
\end{eqnarray}
Here, $\hat{\mu}_{k}^z$ and $\hat{\sigma}_{k}^z$ denote the $z$-component of the standard spin-1/2 operator with two respective eigenvalues $\pm 1/2$ and the coordinate operator $\hat{x}_k$ is assigned to a local displacement of the $k$th decorating atom from its equilibrium lattice position placed in the middle in between its two nearest-neighbour nodal atoms. Note that the displacement $x_k$ will be hereafter 
regarded to be dimensionless (i.e. this parameter determines displacement in multipliers of some characteristic length), which consequently allows one to consider all the other Hamiltonian parameters 
to be in energy units. The exchange constant $J$ determines a magnitude of the nearest-neighbour interaction between decorating and nodal spins provided that the decorating atom takes its equilibrium position, while the magnetoelastic (spin-phonon) coupling constant $A$ describes an appropriate decrease (increase) in the nearest-neighbour exchange interaction invoked by the elongation (contraction) of a relevant distance. 
The elastic part of the bond Hamiltonian $\hat{\cal H}_{k,l}$ includes the kinetic energy 
of the $k$th decorating atom with the mass $M$ and the elastic energy penalty, which is associated 
with the displacement of the $k$th decorating atom from its equilibrium position
\begin{eqnarray}
\hat{{\cal H}}_{k,l} = \frac{\hat{p}_k^2}{2M} + K \hat{x}_k^2.
\label{hal}
\end{eqnarray}
It is noteworthy that the elastic part of bond Hamiltonian given by Eq.~(\ref{hal}) is in fact the Hamiltonian of the quantum harmonic oscillator with twice as large bare elastic (spring stiffness) constant due to a simultaneous distortion of two harmonic springs attached to each decorating atom. 

It should be noted here that an interconnection between spin and lattice degrees of freedom is usually 
the main obstacle, which precludes an exact treatment of the lattice-statistical spin models 
including the spin-phonon interaction. However, our assumption on the incapability of the 
nodal atoms to perform lattice vibrations enables one to decouple spin and lattice degrees of freedom by making use of the local canonical coordinate transformation \cite{enti73}
\begin{eqnarray}
\hat{x}_k = \hat{x}'_k - \frac{A}{2K} {\hat{\mu}}_{k}^z 
            \left({\hat{\sigma}}_{k1}^z - {\hat{\sigma}}_{k2}^z \right).
\label{cct}
\end{eqnarray} 
As a matter of fact, the coordinate transformation (\ref{cct}) eliminates from the magnetoelastic 
part of the bond Hamiltonian (\ref{ham}) dependence on a displacement operator  
\begin{eqnarray}
\hat{{\cal H}}_{k,m}' = - J {\hat{\mu}}_{k}^z \left({\hat{\sigma}}_{k1}^z + {\hat{\sigma}}_{k2}^z \right)
+ \frac{A^2}{8K} {\hat{\sigma}}_{k1}^z {\hat{\sigma}}_{k2}^z - \frac{A^2}{32K},
\label{hamt}
\end{eqnarray}
which is substituted through the effective next-nearest-neighbour interaction between the nodal Ising spins 
and the less important constant term. A presence of the antiferromagnetic next-nearest-neighbour interaction between the nodal spins in the effective spin Hamiltonian (\ref{hamt}) implies an eventual spin frustration, which originates from the magnetoelastic coupling. By contrast, the elastic part of the bond Hamiltonian (\ref{hal}) remains unchanged under the local canonical coordinate transformation (\ref{cct}) 
\begin{eqnarray}
\hat{{\cal H}}_{k,l}' = \frac{(\hat{p}_k')^2}{2M} + K \, (\hat{x}_k')^{2}. 
\label{halt}
\end{eqnarray}
Introducing standard annihilation and creation bosonic operators obeying the commutation 
rule $[\hat{b}_k, \hat{b}_k^{\dagger}] = 1$
\begin{eqnarray}
\hat{b}_k \!=\!\! \sqrt{\frac{M \omega}{2 \hbar}} 
              \left( \hat{x}'_k + \frac{i}{M \omega} \hat{p}'_k \right), \, \, \,
\hat{b}_k^{\dagger} \!=\!\! \sqrt{\frac{M \omega}{2 \hbar}} 
              \left( \hat{x}'_k - \frac{i}{M \omega} \hat{p}'_k \right), \nonumber
\label{ac}
\end{eqnarray}
allows one to bring the elastic part of the bond Hamiltonian (\ref{halt}) 
into the diagonal form 
\begin{eqnarray}
\hat{{\cal H}}_{k,l}' = \hbar \omega \left(\hat{b}_k^{\dagger} \hat{b}_k + \frac{1}{2} \right),
\label{haltac}
\end{eqnarray}
which is more appropriate for further calculations. It is worthy of notice that the angular frequency 
of normal-mode oscillations is given by $\omega=\sqrt{2K/M}$.

Now, let us proceed to a calculation of the partition function. It can be easily understood that the 
total partition function can be partially factorized into a product of bond partition functions, 
because different bond Hamiltonians obviously commute with each other. In addition, the coordinate transformation (\ref{cct}) allows one to treat the elastic and magnetic part of the bond partition function independently of each other and consequently, the partition function can be factorized 
into the following product
\begin{eqnarray}
{\cal Z} \!\!\!&=&\!\!\! 
\sum_{\{ \sigma \}} \prod_{k=1}^{Nq/2} \left[ \mbox{Tr}_{k,l} \exp(- \beta \hat{{\cal H}}_{k,l}')
\!\!\! \sum_{\mu_k^z = \pm 1/2} \!\!\! \exp(- \beta \hat{{\cal H}}_{k,m}') \right] \nonumber \\
\!\!\!&=&\!\!\! \sum_{\{ \sigma \}} \prod_{k=1}^{Nq/2} {\cal Z}_{k,l} {\cal Z}_{k,m}.  
\label{z}
\end{eqnarray}
Here, $\beta = 1/(k_{\rm B} T)$, $k_{\rm B}$ is Boltzmann's constant, $T$ is the absolute temperature,
the symbol $\sum_{\{ \sigma \}}$ marks a summation over all possible configurations of the nodal Ising 
spins and $\mbox{Tr}_{k,l}$ denotes a trace over the lattice degrees of freedom connected to the $k$th decorating atom. The elastic part of the bond partition function ${\cal Z}_{k,l}$ can readily be obtained by employing a trace invariance with respect to unitary transformation and using the diagonalized form of 
the elastic part of the bond Hamiltonian (\ref{haltac})  
\begin{eqnarray}
{\cal Z}_{k,l} = \sum_{n_k=0}^{\infty} \exp \left[-\beta \hbar \omega \left(n_k + \frac{1}{2} \right)\right] 
= \left[2 \sinh \left( \frac{\beta \hbar \omega}{2} \right) \right]^{-1}.
\label{zl}
\end{eqnarray}
Further, the magnetic part of the bond partition function ${\cal Z}_{k,m}$ can be replaced with a simpler equivalent expression provided by the generalized decoration-iteration transformation \cite{fish59,roja09,stre10,lap}
\begin{eqnarray}
{\cal Z}_{k,m} \!\!\!\!&=&\!\!\!\! 2 \exp \left[ \frac{\beta A^2}{32K} \left(1 -  4 \sigma_{k1}^z \sigma_{k2}^z \right) \right] \cosh \left[\frac{\beta J}{2} (\sigma_{k1}^z + \sigma_{k2}^z) \right] \nonumber \\ \!\!\!\!&=&\!\!\!\! R_0 \exp(\beta R_1 \sigma_{k1}^z \sigma_{k2}^z),
\label{dit}
\end{eqnarray}
which is of a general validity if and only if the mapping parameters $R_0$ and $R_1$ 
are given by the following formulas
\begin{eqnarray}
R_0 \!\!\!&=&\!\!\! 2 \exp \left(\frac{\beta A^2}{32K} \right) 
\sqrt{\cosh \left(\frac{\beta J}{2} \right)}, \label{mp0} \\
\beta R_1 \!\!\!&=&\!\!\! - \frac{\beta A^2}{8K} + 2 \ln \left[ \cosh \left(\frac{\beta J}{2} \right) \right].
\label{mp1}
\end{eqnarray}
Substituting the elastic part of the bond partition function (\ref{zl}) together with 
the generalized decoration-iteration transformation (\ref{dit}) into the formula (\ref{z}), one readily 
gains an exact mapping relationship between the partition function of the spin-1/2 Ising model
on a planar lattice with the decorating atoms amenable to lattice vibrations and respectively, 
the partition function of the spin-1/2 Ising model on a corresponding undecorated rigid lattice 
\begin{eqnarray}
{\cal Z} = \left[ \frac{\exp \left(\frac{\beta A^2}{16K} \right) \cosh \left(\frac{\beta J}{2} \right)}
           {\sinh^2 \left( \frac{\beta \hbar \omega}{2} \right)} \right]^{\frac{Nq}{4}} 
           {\cal Z}_{\rm IM} (\beta, R_1).
\label{mr}
\end{eqnarray}
Thus, the partition function of the investigated model system can readily be obtained from the known exact results \cite{onsa44,hout50,domb60,mcoy73,lavi99} for the partition function of the spin-1/2 Ising model on undecorated planar lattices with the effective interaction $R_1$ given by Eq.~(\ref{mp1}). Besides, the critical behaviour of the spin-1/2 Ising model on a decorated lattice with the vibrating decorating atoms can easily be examined when comparing the effective nearest-neighbour coupling (\ref{mp1}) of the corresponding 
spin-1/2 Ising model on undecorated rigid lattice with its critical value (e.g. $\beta_c |R_1| = 2 \ln (1 + \sqrt{2})$ for a square lattice \cite{onsa44}). It is also worth mentioning that the established mapping relationship (\ref{mr}) between both partition functions permits a straightforward calculation of other thermodynamic quantities such as the internal energy 
\begin{eqnarray}
{\cal U} \!\!\!&=&\!\!\! {\cal U}_{L} + {\cal U}_M = 
\frac{Nq}{4} \hbar \omega \, \mbox{coth} \left( \frac{\beta \hbar \omega}{2} \right)
\label{u} \\
\!\!\!&-&\!\!\! \frac{Nq}{4} \left[\frac{A^2}{16 K} (1 - 4 \varepsilon_{\rm IM}) 
+ \frac{J}{2} \tanh \left(\frac{\beta J}{2} \right) (1 + 4 \varepsilon_{\rm IM}) \right]. \nonumber
\end{eqnarray}
Here, the former (latter) expression represents the lattice (magnetic) part of the overall internal 
energy and the quantity $\varepsilon_{\rm IM} \equiv \langle \hat{\sigma}_{k1}^z 
\hat{\sigma}_{k2}^z \rangle$ labels the nearest-neighbour pair correlation function of the corresponding 
spin-1/2 Ising model on undecorated (rigid) lattice \cite{domb60,mcoy73,lavi99}. Note that the specific heat can be obtained from Eq.~(\ref{u}) by the use of well-known relation $C = \partial {\cal U}/\partial T$, but the final formula is too cumbersome to write it down here explicitly.

Contrary to this, the spontaneous magnetization cannot be simply obtained from the mapping relation
(\ref{mr}) between the partition functions. The uniform and staggered magnetizations of the nodal Ising spins can be however derived by employing exact mapping theorems developed by Barry \textit{et al}. \cite{barr88,khat90,barr91,barr95}, which establish a precise mapping equivalence between the ensemble 
average $\langle \cdots \rangle$ in the spin-1/2 Ising model on a decorated lattice and the 
ensemble average $\langle \cdots \rangle_{\rm IM}$ in the corresponding spin-1/2 Ising model on 
the undecorated rigid lattice for any function depending just on the nodal Ising spins. Accordingly, 
the uniform and staggered magnetizations of the nodal Ising spins can be obtained from the mapping relations
\begin{eqnarray}
\!\!\!\!\!\!&&\!\!\!\!\!\! m_A \!\equiv\! \frac{1}{2} \!
\left \langle \hat{\sigma}_{k1}^z \!\!+\! \hat{\sigma}_{k2}^z \right \rangle
\!=\! \frac{1}{2} \! \left \langle  \hat{\sigma}_{k1}^z \!\!+\! \hat{\sigma}_{k2}^z \right \rangle_{\rm IM} 
\!\! \equiv \! m_{\rm IM} (\beta R_1\!>\!0)\!, \label{ma}  \\ 
\!\!\!\!\!\!&&\!\!\!\!\!\! s_A \!\equiv\! \frac{1}{2} \!
\left \langle \hat{\sigma}_{k1}^z \!\!-\! \hat{\sigma}_{k2}^z \right \rangle
\!=\! \frac{1}{2} \! \left \langle \hat{\sigma}_{k1}^z \!\!-\! \hat{\sigma}_{k2}^z \right \rangle_{\rm IM} 
\!\! \equiv \! s_{\rm IM} (\beta R_1\!<\!0), \label{sa} 
\end{eqnarray} 
which connect both these quantities with the spontaneous (either uniform or staggered) magnetization 
of the corresponding spin-1/2 Ising model on the undecorated rigid lattice \cite{yang52,lin92}. On the other hand, the spontaneous magnetization of the decorating Ising spins $m_B$ can be related to the spontaneous magnetization of the nodal Ising spins $m_A$ with the aid of the generalized Callen-Suzuki identity \cite{call63,suzu65,balc02} 
\begin{eqnarray}
m_B \equiv \! \left \langle \hat{\mu}_{k}^z  \right \rangle \! = \left \langle \frac{1}{2} 
\tanh \left[ \frac{\beta J}{2} (\hat{\sigma}_{k1}^z + \hat{\sigma}_{k2}^z) \right] \right \rangle \!
 = m_A \tanh \left( \frac{\beta J}{2} \right). \nonumber
\label{mb} 
\end{eqnarray} 

Last, let us calculate the mean displacement and standard deviation of the decorating atoms from their equilibrium lattice positions. It is quite obvious from the coordinate transformation (\ref{cct}) that the mean displacement can be expressed in terms of two spin correlation functions and the mean displacement 
in a new coordinate system   
\begin{eqnarray}
\langle \hat{x}_k \rangle = \langle \hat{x}'_k \rangle - \frac{A}{2K} \left( \langle {\hat{\mu}}_{k}^z {\hat{\sigma}}_{k1}^z \rangle - \langle {\hat{\mu}}_{k}^z {\hat{\sigma}}_{k2}^z \rangle \right).
\label{md}
\end{eqnarray}  
It can be readily proved that both spin correlation functions between the decorating spin 
and its two nearest-neighbour nodal spins are identical
\begin{eqnarray}
\langle {\hat{\mu}}_{k}^z {\hat{\sigma}}_{k1}^z \rangle = 
\langle {\hat{\mu}}_{k}^z {\hat{\sigma}}_{k2}^z \rangle =
\frac{1}{8} (1 + 4 \varepsilon_{\rm IM}) \tanh \left( \frac{\beta J}{2} \right) 
\label{scf}
\end{eqnarray} 
and the mean displacement in a shifted coordinate system equals zero
\begin{eqnarray}
\langle \hat{x}'_k \rangle = \sqrt{\frac{\hbar}{2 M \omega}} 
(\langle \hat{b}_{k}^{\dagger} \rangle + \langle \hat{b}_{k} \rangle) = 0.
\label{mdis} 
\end{eqnarray} 
These results are taken to mean that the mean displacement also equals zero, i.e. $\langle \hat{x}_k \rangle = 0$, and the decorating atoms oscillate symmetrically around their equilibrium lattice positions. To bring 
an insight into a magnitude of those lattice oscillations, let us also calculate the standard deviation 
for the displacement of the decorating atoms from their equilibrium lattice positions using
\begin{eqnarray}
d \equiv \sqrt{\langle \hat{x}_k^2 \rangle - \langle \hat{x}_k \rangle^2} = 
\sqrt{\langle (\hat{x}'_k)^2 \rangle + \frac{A^2}{32K^2} \left( 1 - 4 \varepsilon_{\rm IM} \right)}.
\label{rsm}
\end{eqnarray}  
The mean of a square of the displacement of the decorating atoms in a shifted 
coordinate system is given by 
\begin{eqnarray}
\langle (\hat{x}'_k)^2 \rangle = \frac{\hbar}{M \omega} \left( \left \langle \hat{b}_k^{\dagger} \hat{b}_k \right \rangle + \frac{1}{2} \right) = 
\frac{\hbar \omega}{4 K} \, \mbox{coth} \left( \frac{\beta \hbar \omega}{2} \right)
\label{rsmn}
\end{eqnarray}
and hence, the standard deviation for the displacement of the decorating atoms 
readily follows from the formula
\begin{eqnarray}
d = \sqrt{\frac{\hbar \omega}{4 K} \, \mbox{coth} \left( \frac{\beta \hbar \omega}{2} \right) 
         + \frac{A^2}{32K^2} \left( 1 - 4 \varepsilon_{\rm IM} \right)}. 
\label{sd}
\end{eqnarray} 
 
\section{Results and discussion}
\label{result}

Now, let us proceed to an analysis of the most interesting results obtained in the foregoing section.
Although all generic features of the investigated model system will be hereafter illustrated only 
on a particular example of the spin-1/2 Ising model on a decorated square lattice with the coordination 
number $q=4$, the qualitatively same behaviour may be expected for any other decorated loose-packed 
planar lattice as well. It is noteworthy, moreover, that the exchange and magnetoelastic constants can be considered without loss of generality positive, because the relevant change in a character of the exchange interaction does not fundamentally affect neither the critical behaviour nor the behaviour of basic thermodynamic quantities as it merely causes a rather trivial change in a relative orientation of the decorating and nodal Ising spins, respectively. Last but not least, let us introduce the following set of reduced parameters $A/J$, $K/J$, $\hbar \omega/J$ and $k_{\rm B} T/J$, which measure a relative strength of the magnetoelastic constant, the spring stiffness constant, the phonon energy and the temperature, respectively.

\begin{figure}
\begin{center}
\includegraphics[width=0.4\textwidth]{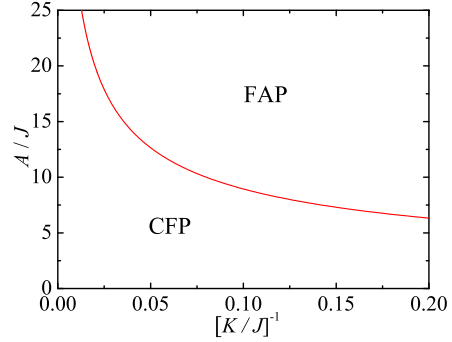}
\end{center}
\vspace{-1.1cm}
\caption{The ground-state phase diagram in the $J/K - A/J$ plane.}
\label{fig2}       
\end{figure}

The typical ground-state phase diagram is depicted in Fig.~\ref{fig2} in the the $J/K - A/J$ plane. 
It turns out that the classical ferromagnetic phase (CFP) with a perfect parallel alignment of all 
nodal as well as decorating spins constitutes the ground state when the magnetoelastic coupling is smaller than the boundary value $A < A_{\rm b} = \sqrt{8 K J}$. On the other hand, the more striking frustrated antiferromagnetic phase (FAP) becomes the ground state whenever the magnetoelastic coupling exceeds the boundary value $A>A_{\rm b}$. 
The spin arrangement inherent to FAP can be characterized by a remarkable coexistence of a partial order
and disorder, since the perfect antiferromagnetic (N\'eel) long-range order of the nodal spins 
is accompanied with a complete disorder of the decorating spins. An origin of this 
rather unexpected and unusual phase apparently lies in the spin-phonon coupling induced spin frustration, which manifest itself in the competing character of the next-nearest-neighbour interaction between the 
nodal spins in the effective spin Hamiltonian (\ref{hamt}) and leads to a frustration of 
the decorating spins.

\begin{figure}
\begin{center}
\includegraphics[width=0.4\textwidth]{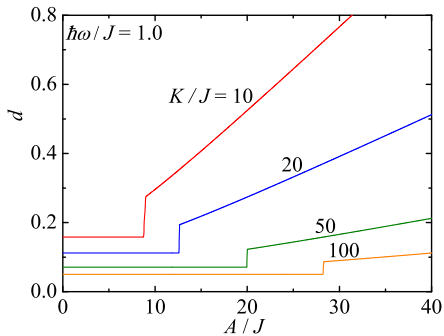}
\end{center}
\vspace{-1.1cm}
\caption{The zero-temperature dependence of the standard deviation 
on the magnetoelastic constant for several values of the spring stiffness
constant and the fixed phonon energy $\hbar \omega/J = 1$.}
\label{fig3}       
\end{figure}

To bring an insight into how the phase transition between CFP and FAP manifest itself in the lattice vibrations of the decorating atoms, Fig.~\ref{fig3} displays the zero-temperature dependence of 
the standard deviation on a relative strength of the magnetoelastic constant. Even though the decorating 
atoms oscillate symmetrically around their equilibrium lattice positions both in CFP and FAP,
it is quite evident from Fig.~\ref{fig3} that the magnitude of these lattice oscillations are 
much more robust in FAP than CFP. As a matter of fact, the standard deviation does 
not depend on the magnetoelastic constant within the region corresponding to CFP where
\begin{eqnarray}
d_{\rm CFP} = \sqrt{\frac{\hbar \omega}{4 K}}, 
\label{sdcfp}
\end{eqnarray} 
while it jumps towards a higher value at $A=A_{\rm b}$ and then increases monotonically with 
the magnetoelastic constant within the region corresponding to FAP where
\begin{eqnarray}
d_{\rm FAP} = \sqrt{\frac{\hbar \omega}{4 K} + \frac{A^2}{16K^2}}. 
\label{sdfap}
\end{eqnarray}
From this point of view, our further analysis will be restricted only to a discussion of the particular cases with a sufficiently small value of the ratio $A/K$ between the magnetoelastic and spring stiffness constants in order to keep a magnitude of lattice oscillations within the range of validity of 
the harmonic approximation.

\begin{figure}
\begin{center}
\includegraphics[width=0.4\textwidth]{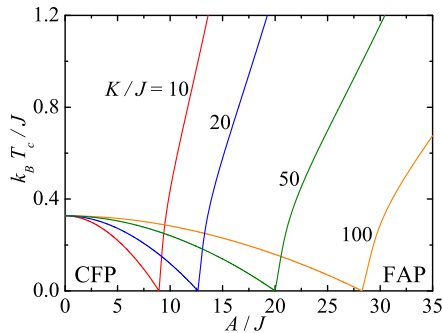}
\end{center}
\vspace{-1.1cm}
\caption{The critical temperature as a function of the magnetoelastic coupling
for several values of the spring stiffness constant.}
\label{fig4}       
\end{figure}

Now, let us proceed to a discussion of the critical behaviour. In Fig.~\ref{fig4}, the critical 
temperature is plotted against a relative strength of the magnetoelastic constant for several values 
of the spring stiffness constant. The critical temperature of CFP gradually decreases by increasing 
the magnetoelastic coupling until it completely vanishes at the ground-atate boundary between CFP and FAP.
On the contrary, one may observe a rather steep increase of the critical temperature upon further increase 
of the magnetoelastic coupling in FAP, which can be mainly attributed to a quadratic dependence of the antiferromagnetic part of the effective interaction (\ref{mp1}) on the magnetoelastic constant. It is worthwhile to recall, however, that the validity of our results for the critical temperatures of FAP 
is restricted only to a rather narrow region in a close vicinity of the ground-state phase boundary between CFP and FAP, where the standard deviation is small enough to ensure the validity of harmonic approximation.

\begin{figure}
\begin{center}
\includegraphics[width=0.4\textwidth]{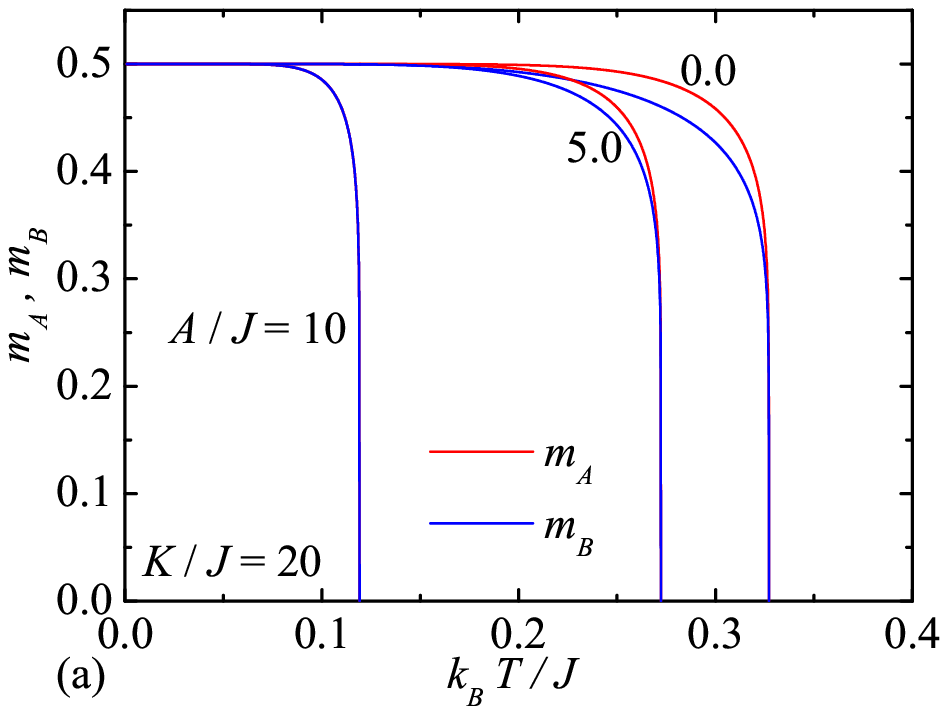}
\includegraphics[width=0.4\textwidth]{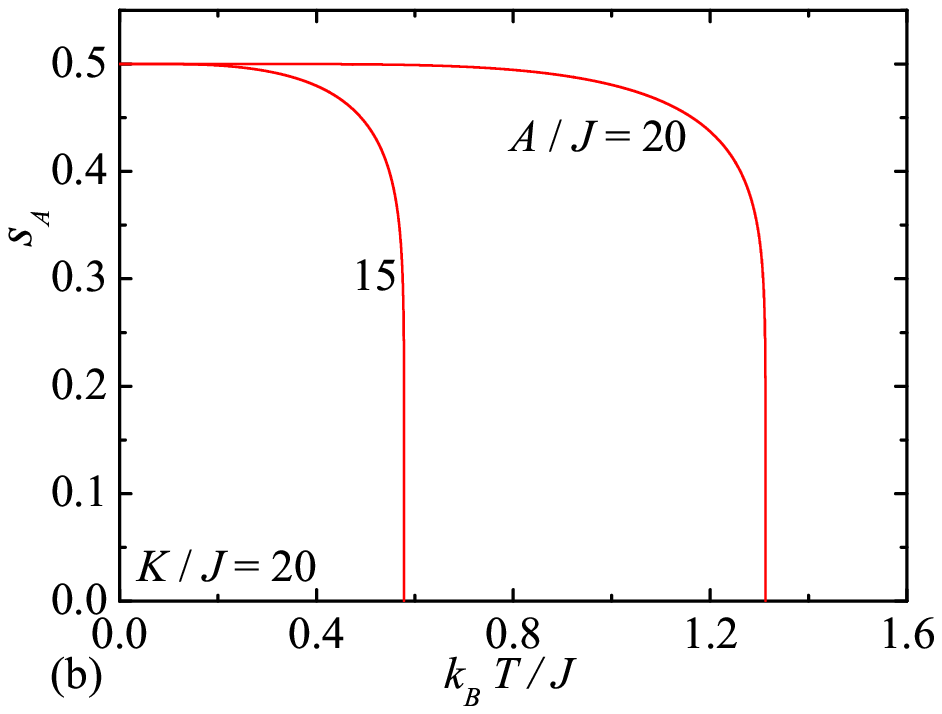}
\end{center}
\vspace{-1.1cm}
\caption{(a) Thermal dependence of both spontaneous sublattice magnetizations for the fixed value of the spring stiffness constant $K/J = 20$ and three different values of the magnetoelastic constant;
(b) Thermal dependence of the spontaneous staggered magnetization of the nodal Ising spins 
for the fixed value of the spring stiffness constant $K/J = 20$
and two different values of the magnetoelastic constant.}
\label{fig5}
\end{figure}

Next, let us turn our attention to temperature dependences of the spontaneous magnetization. 
Fig.~\ref{fig5}(a) shows typical thermal variations of both spontaneous sublattice magnetizations  
in CFP for three different values of the magnetoelastic constant. The sublattice magnetization $m_A$ of 
the nodal Ising spins is always more resistent to thermal fluctuations than the sublattice
magnetization $m_B$ of the decorating Ising spins even if both sublattice magnetizations tend 
to zero with the same critical exponent $\beta_e=1/8$ from the standard Ising universality class. 
It is worthy to notice, moreover, that the difference in the thermal behaviour of both sublattice
magnetizations is the smaller, the stronger is a relative strength of the magnetoelastic constant.
On the other hand, the relevant order parameter for partially ordered and partially disordered 
FAP represents the staggered magnetization $s_A$ of the nodal Ising spins, while the decorating 
Ising spins remain disordered over the whole temperature range in FAP. The typical temperature 
variations of the staggered magnetization of the nodal Ising spins are depicted in Fig.~\ref{fig5}(b). 
It is worth mentioning that the thermal behaviour of the staggered magnetization in a close vicinity 
of the critical point is also governed by the critical exponent $\beta_e=1/8$ from the standard 
Ising universality class.

\begin{figure}
\begin{center}
\includegraphics[width=0.4\textwidth]{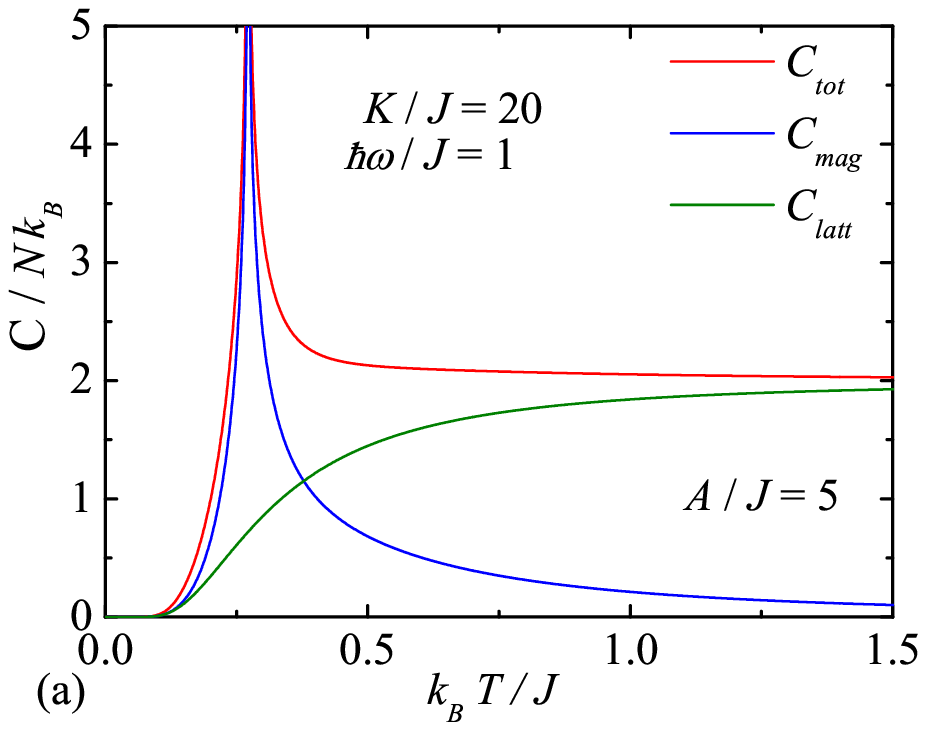}
\includegraphics[width=0.4\textwidth]{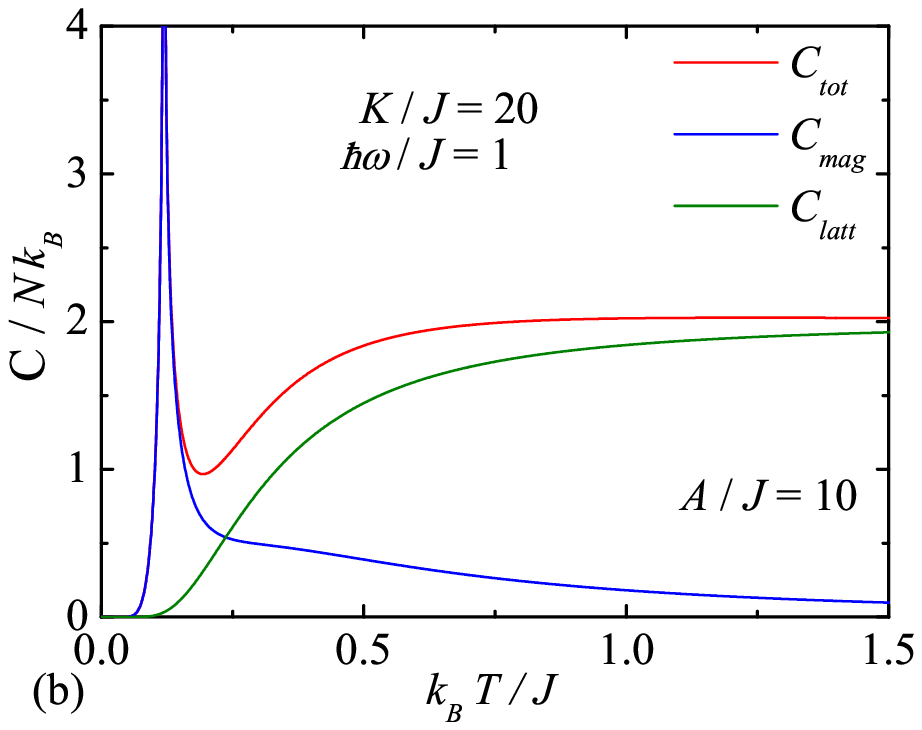}
\end{center}
\vspace{-1.1cm}
\caption{Temperature dependences of the total specific heat, its magnetic and lattice parts 
for the fixed value of the spring stiffness constant $K/J = 20$, the phonon energy $\hbar \omega/J = 1$ 
and two different values of the magnetoelastic constant: (a) $A/J = 5$, 
(b) $A/J = 10$.}
\label{fig6}
\end{figure}

\begin{figure}
\begin{center}
\includegraphics[width=0.4\textwidth]{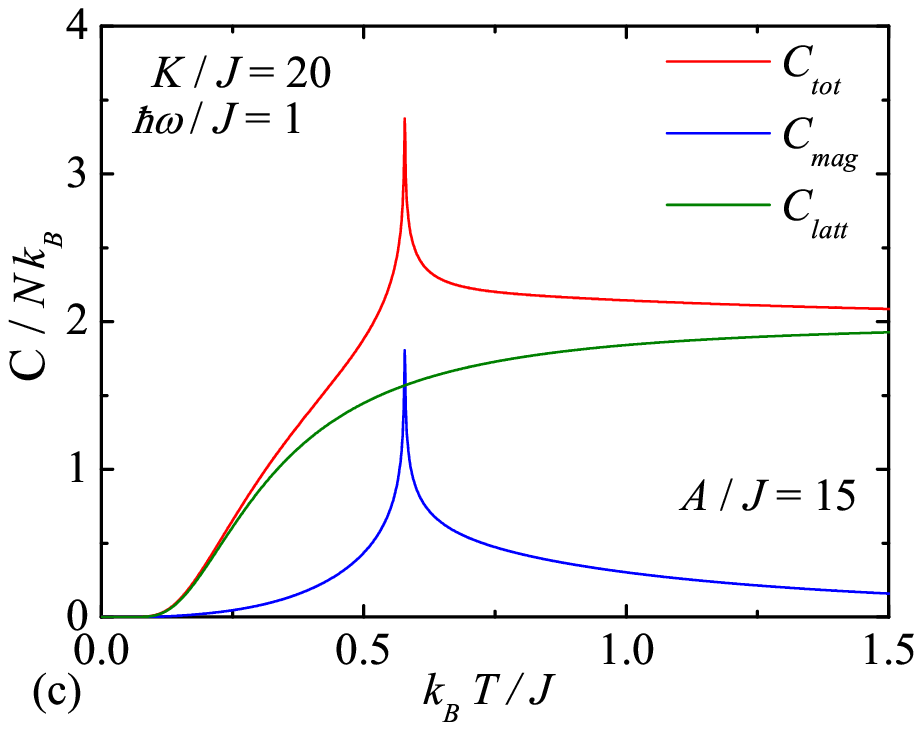}
\includegraphics[width=0.4\textwidth]{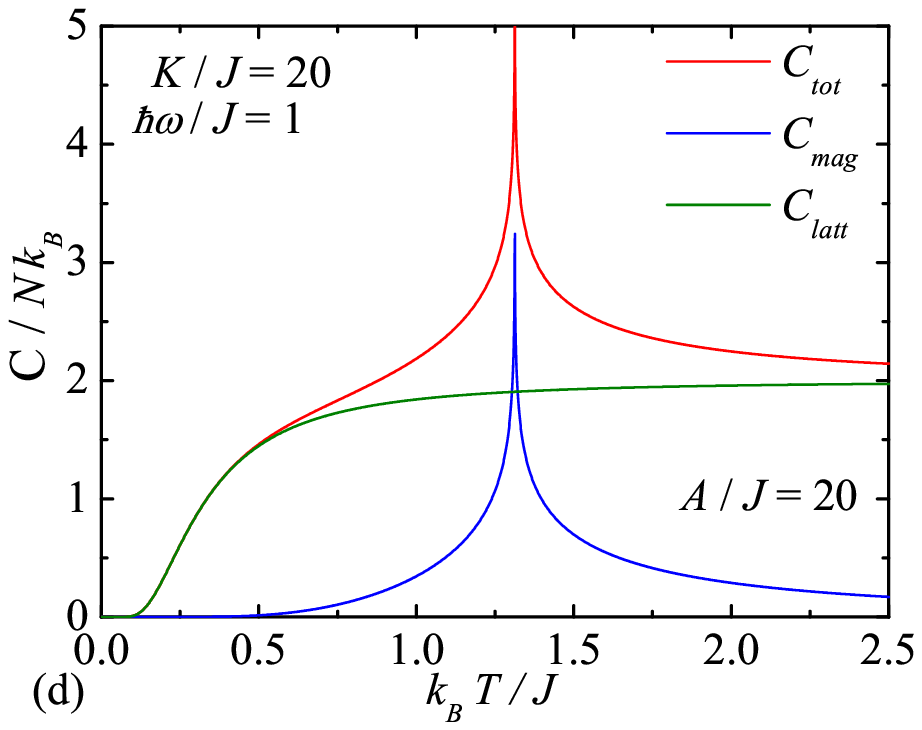}
\end{center}
\vspace{-1.1cm}
\caption{The same figure caption as for Fig.~\ref{fig6}, but for the magnetoelastic constant: 
(a) $A/J = 15$, (b) $A/J = 20$.}
\label{fig7}
\end{figure}

Last but not least, let us discuss the main features that are typical for temperature variations of the specific heat. Fig.~\ref{fig6} illustrates two typical thermal dependences of the specific heat for CFP, 
while Fig.~\ref{fig7} depicts another two typical temperature dependences of the specific heat for FAP. 
To gain a deeper insight into the overall behaviour, both lattice as well as magnetic part of the specific heat are plotted along with the total specific heat in those figures. It is quite obvious from Fig.~\ref{fig6}(a) that the temperature variations of the specific heat are initially governed almost exclusively by its magnetic contribution showing 
the $\lambda$-type singularity at the critical temperature, while the phonon contribution is just 
superimposed on the high-temperature tail of the magnetic heat capacity if considering relatively 
small values of the magnetoelastic constant. Another interesting non-monotonic thermal dependence 
of the specific heat with the $\lambda$-type singularity superimposed on a round shoulder can be 
detected in Fig.~\ref{fig6}(b), where the magnetic and phonon contributions to the overall heat 
capacity are even better separated one from each other due to a relevant decrease in the critical 
temperature caused by an enhancement of the magnetoelastic coupling. This leads to an appearance of 
the round minimum in the high-temperature tail of the $\lambda$-type singularity before the specific 
heat finally approaches its constant value $C = 2 N k_{\rm B}$ in the limit of sufficiently high temperatures.

Finally, Fig.~\ref{fig7}(a)-(b) illustrate typical thermal dependences of the specific heat in FAP, 
where a substantial shift in the critical temperature can be achieved by a rather small increase 
in a relative strength of the magnetoelastic constant. In this particular case, the lattice 
contribution to the overall heat capacity may prevail over the magnetic contribution even 
in the low-temperature region and hence, the $\lambda$-type singularity can be superimposed 
either on the ascending part (Fig.~\ref{fig7}(a)) or the saturated part (Fig.~\ref{fig7}(b)) of 
the lattice specific heat. Thus, the round shoulder in the low-temperature tail of the specific heat 
observed in Fig.~\ref{fig7}(b) obviosly comes from the phonon rather than magnetic excitations.
It is worthwhile to remark, moreover, that a few other types of dependences mostly with a rather 
well separated magnetic and lattice contributions can be achieved by varying the phonon energy 
and the spring stiffness constant, however, our particular attention was mainly concentrated 
on special cases with an obvious interplay between both contributions.

\section{Concluding remarks}
\label{conclusion}

The present Letter deals with the magnetoelastic properties of the spin-1/2 Ising model with 
a spin-phonon coupling on decorated planar lattices partially prone to lattice vibrations, 
which are investigated by making use of the generalized decoration-iteration transformation 
and the harmonic approximation. Within the framework of this approach, we have examined 
the ground-state and finite-temeprature phase diagrams, the mean and standard deviation 
of the displacement of the decorating atoms from their equilibrium lattice positions, 
as well as, the temperature dependences of the spontaneous magnetization and the specific heat.

It has been demonstrated that the spin-phonon interaction gives rise to an effective antiferromagnetic next-nearest-neighbour interaction between the nodal spins, which generally competes with the nearest-neighbour interaction between the decorating and nodal spins. As a result of this competition, 
one may encounter for sufficiently strong magnetoelastic couplings a quite intriguing partially ordered 
and partially disordered phase, where the antiferromagnetic long-range order of the nodal spins 
is accompanied with a complete disorder of the decorating spins. 

Before concluding, it is worthwhile to remark that the approach presented in this Letter 
can also be rather straightforwardly generalized to the mixed spin-(1/2, $S$) Ising model 
on decorated planar lattices with the decorating atoms of a quite general spin $S \geq 1$, 
where the spin-phonon coupling will produce an effective single-ion anisotropy and three-site 
four-spin interaction. Our future goal is to clarify the effect of the magnetoelastic coupling 
on a magnetic behaviour of these more general mixed-spin Ising models.

\end{document}